\documentclass{ws-p10x7}

\begin{document}

\title{Improved SUSY QCD corrections to Higgs boson decays 
into quarks and squarks}

\author{Y. Yamada$^1$, H. Eberl$^2$, K. Hidaka$^3$, S. Kraml$^2$, 
W. Majerotto$^2$}

\address{$^1$ Department of Physics, Tohoku University, Sendai 980--8578, 
Japan }

\address{$^2$ Institut f\"ur Hochenergiephysik 
der \"Osterreichischen Akademie der Wissenschaften,\\A--1050 Vienna, 
Austria }

\address{$^3$ Department of Physics, Tokyo Gakugei University, Koganei,
Tokyo 184--8501, Japan }

\twocolumn[\maketitle\abstract{
The ${\cal O}(\alpha_s)$ SUSY QCD corrections to the decays of 
the MSSM Higgs bosons into bottom quarks and squarks can be very 
large for large $\tan\beta$ in the on--shell renormalization scheme. 
We improve the calculation by a careful choice of the tree-level 
Higgs boson couplings in terms of running parameters of quarks 
and squarks. 
}]

\section{Introduction}

The MSSM has two Higgs doublets ($H_1$, $H_2$) which give five physical 
bosons ($h^0$, $H^0$, $A^0$, $H^\pm$). 
Their couplings to bottom quarks $b$ and squarks $\tilde{b}$ 
are enhanced for large $\tan\beta$. In this case the decays to $b$ are 
usually the main modes.\cite{review} Decays to $\tilde{b}$ can be also 
dominant.\cite{hsqtree} Studying these decays is therefore very important. 

These decays receive large SUSY QCD corrections.\cite{hqq,hsq} When the 
on--shell scheme is adopted for quarks and squarks, 
the corrections are often very large and make the 
perturbation calculation quite unreliable. 
The large gluon loop correction can be absorbed by using the 
QCD running quark mass in the coupling. 
However, the gluino loop correction can also 
be very large for large $\tan\beta$. 

Here we improve\cite{mainarticle} the one--loop SUSY QCD 
corrected widths of the Higgs boson decays into quarks and squarks. 
The essential point of the improvement is to define 
appropriate tree--level couplings of the Higgs bosons 
to $b$ and $\tilde{b}$. 

\section{Gluino corrections to Higgs--quark couplings}

The main part of the \pagebreak large gluino loop corrections to the 
Higgs decay widths into $b$ originates from the $\bar{b}bH_2$ coupling 
which is generated by squark--gluino loops. 

At tree--level, the $\bar{b}bH_2$ coupling is forbidden by SUSY. 
However, the interaction $h_b\Delta_b\bar{b}bH_2$ is generated by the 
loop correction due to the soft SUSY breaking. 
The squark-gluino loops give 
$\Delta_b\sim\alpha_s m_{\tilde{g}}\mu/m_{\tilde{b}}^2$. $\Delta_b$ can 
have further contributions from other loop 
corrections.\cite{dmb,hbbnew} 

The effective interactions between Higgs bosons and $b$, after 
integrating out the squarks, are properly described by 
\begin{eqnarray}
{\cal L}_{\rm int}^{\rm eff}&=& 
-(h_b/\sqrt{2})\bar{v}[\cos\beta+\Delta_b \sin\beta]\,
\bar{b}b \nonumber \\
&&-(h_b/\sqrt{2})[\cos\alpha+\Delta_b \sin\alpha]\,H^0\bar{b}b 
\nonumber \\
&&+(h_b/\sqrt{2})[\sin\alpha-\Delta_b \cos\alpha]\,h^0\bar{b}b 
\nonumber \\
&&+(ih_b/\sqrt{2})[\sin\beta-\Delta_b \cos\beta]\,A^0\bar{b}\gamma_5b 
\nonumber \\
&&+h_b[\sin\beta-\Delta_b \cos\beta]\,H^-\bar{b}_R t_L +({\rm h.c.}).
\nonumber \\
\label{e13}
\end{eqnarray}
The first term of Eq.~(\ref{e13}) gives the (non--SUSY) QCD 
running mass $m_b(Q)_{\rm SM}$. 
The difference from the SUSY QCD running mass 
$m_b(Q)_{\rm MSSM}=(h_b/\sqrt{2})\bar{v}\cos\beta$ is enhanced by 
$\tan\beta$. As a result, the gluino loop correction to $m_b$ 
can become very large\cite{dmb} for large $\tan\beta$. 

In Eq.~(\ref{e13}) the contributions of $\Delta_b$ to 
the Higgs--bottom couplings take forms different from those to $m_b$. 
When the tree--level couplings are given in terms of $m_b(Q)_{\rm SM}$ or 
the on--shell mass $M_b$, the corrections by $\Delta_b$ can be enhanced 
very much\cite{hbbnew,mainarticle,htbnew} for $\tan\beta\gg 1$. 
This is the main source of the large gluino loop 
corrections to decay widths to $b$ in the on--shell scheme. 
Note that $\Delta_b$ itself is smaller than one and therefore 
does not destroy the validity of the perturbation expansion. 

We can improve the QCD perturbative expansion by changing 
the choice of the tree--level Higgs-bottom couplings. 
For example, when the tree--level $A^0\bar{b}b$ coupling is 
expressed in terms of $m_b(Q)_{\rm MSSM}$ at $Q=m_A$, 
the correction from $\Delta_b$ becomes very small. 
We therefore expect that $m_b(Q)_{\rm MSSM}$ is an 
appropriate parameter for the $A^0\rightarrow\bar{b}b$ decay. This is 
also the case for the $H^+\rightarrow t\bar{b}$ decay. 
The $H^0$ and $h^0$ decays need a special treatment. 
For very large $m_A$, the $H^0\bar{b}b$ and $h^0\bar{b}b$ couplings 
are properly parametrized by $m_b(Q)_{\rm MSSM}$ and $m_b(Q)_{\rm SM}$, 
respectively. In general, the appropriate 
tree--level couplings are given by their linear combinations. 

\section{Higgs--squark couplings}

The large SUSY QCD corrections to the Higgs decays into squarks 
in the on--shell scheme\cite{hsq} mainly come from 
the counterterms for the Higgs--squark couplings, which 
depend on ($m_q$, $\theta_{\tilde{q}}$, $A_q$). 
As in the decays to quarks, we can improve the perturbation 
calculation by using SUSY QCD running parameters $m_q(Q)_{\rm MSSM}$ 
and $A_q(Q)$ in the tree--level couplings. 
However, the mixing angles $\theta_{\tilde{q}}$ are 
kept on--shell in order to cancel the $\tilde{q}_1-\tilde{q}_2$ mixing 
squark wave function corrections. 

\section{Numerical results}

We calculated\cite{mainarticle} the one--loop SUSY QCD corrected 
widths of the Higgs boson decays to $b$ and $\tilde{b}$, 
with and without the improvement presented here. 
In obtaining $m_b(Q)_{\rm MSSM}$ from $m_b(Q)_{\rm SM}$, we express 
the sbottom parameters in the sbottom--gluino loops in terms 
of $m_b(Q)_{\rm MSSM}$ and perform an iteration procedure. 
The large higher-order gluino corrections to $m_b$ are 
then resummed.\cite{mainarticle,htbnew}

\begin{figure}
\epsfxsize200pt
\figurebox{}{}{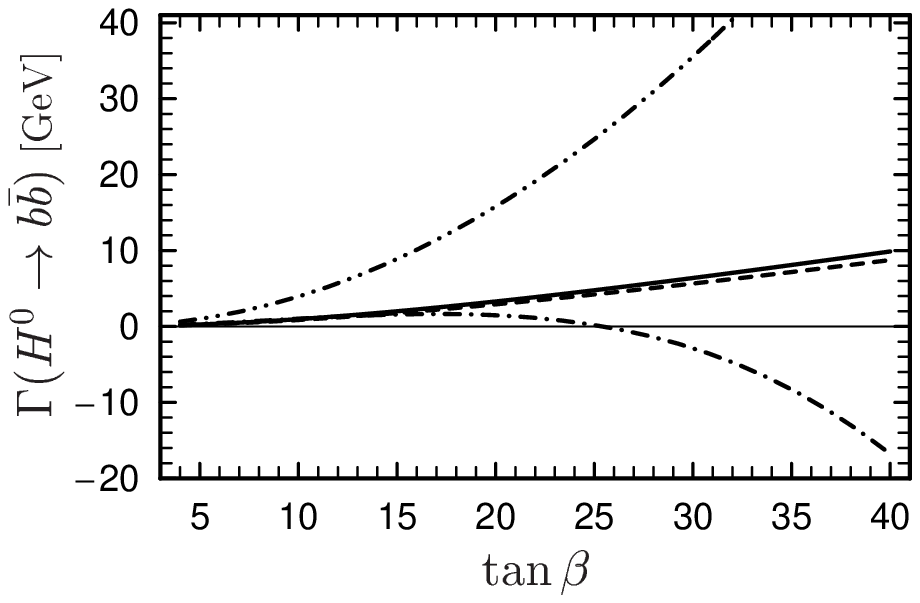}
\caption{Decay width of $H^0\rightarrow b\bar{b}$ 
as a function of $\tan\beta$.
Dash--dot--dotted, dash--dotted, dashed, and full lines 
correspond to the on--shell tree--level, on--shell one--loop, 
improved tree--level, and improved one--loop results, respectively. 
The SUSY parameters are 
$(M_{\tilde{Q}},M_{\tilde{U}},M_{\tilde{D}})=(300,270,330)$ GeV, 
$A_t=150$ GeV, $A_b(Q=m_A)=-700$ GeV, 
$(m_{\tilde{g}},\mu,m_A)=(350,260,800)$~GeV. 
}
\label{fig:1}
\end{figure}

\begin{figure}
\epsfxsize200pt
\figurebox{}{}{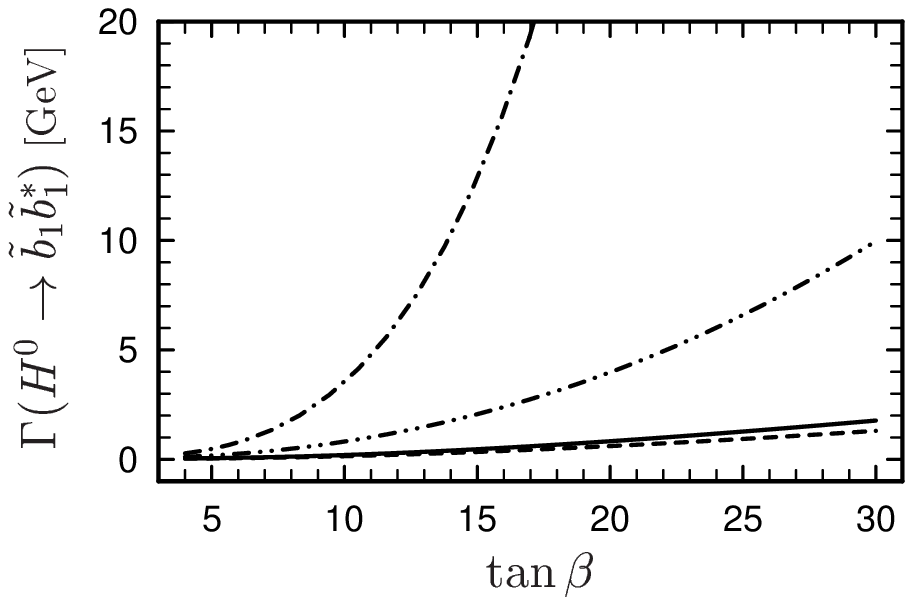}
\caption{Decay width of $H^0\rightarrow\tilde{b}_1\tilde{b}_1^*$ 
as a function of $\tan\beta$. Notations and parameters are 
the same as in Fig.~1. 
}
\label{fig:2}
\end{figure}

Here we show the tree--level and corrected widths of the 
decay $H^0\rightarrow b\bar{b}$ in Fig.~\ref{fig:1}, 
and those of the decay $H^0\rightarrow\tilde{b}_1\tilde{b}_1^*$ 
in Fig.~\ref{fig:2}. 
One can clearly see that the differences between 
tree--level and corrected widths decrease dramatically 
by our method, demonstrating the improvement of the 
perturbation expansion. 

\section{Summary} 

We have improved the SUSY QCD corrections to the Higgs 
decays into $b$ and $\tilde{b}$. 
The essential point of the improvement is to define appropriate 
tree--level couplings of the Higgs bosons to $b$ and $\tilde{b}$, 
in terms of the running parameters of quarks and squarks. 
We have also shown the numerical improvement of the SUSY QCD corrected 
decay widths. 

We note that our method will also be useful in studying other processes 
with Higgs bosons. 

\section*{Acknowledgments}
The work of Y.\,Y. was supported in part by the Grant--in--aid 
for Scientific Research from the Ministry of Education, Science, 
Sports, and Culture of Japan, No.~10740106. 
H.\,E., S.\,K., and W.\,M. thank the 
``Fonds zur F\"orderung der wissenschaftlichen Forschung of Austria'', 
project no. P13139-PHY for financial support.

\end{document}